\documentclass[aip,jcp,reprint,longbibliography]{revtex4-1} 
\usepackage[utf8]{inputenc}

\usepackage{cmap} 
\usepackage[T1]{fontenc} 

\usepackage{amsmath,amssymb,mathtools}
\IfFileExists{mtpro2.sty}
{ 
  \usepackage{times}
  \usepackage[lite,eucal,subscriptcorrection,slantedGreek,zswash]{mtpro2}
}{
\usepackage{txfonts,eucal}

}

\newcommand\dbltilde[1]{\overset{\hbox{\tiny$\approx$}}{#1}}

\usepackage{mleftright}
\usepackage{microtype}

\usepackage{graphicx,grffile,subfigure}
\graphicspath{{figures/}}
\newlength{\figwidth}
\usepackage{url,hyperref}
\usepackage[table,usenames,dvipsnames]{xcolor}
\hypersetup{colorlinks=true, linkcolor=BrickRed,
  urlcolor=blue!50!black, citecolor=blue!50!black}

\usepackage[capitalize]{cleveref}
\crefrangelabelformat{equation}{(#3#1#4)$-$(#5#2#6)}

\usepackage{textcomp}
\usepackage{braket} 

\usepackage{tikz} %

\newcommand{\kb}{k_\text{B}}
\renewcommand{\vec}[1]{\mathbf{#1}}

\newcommand{\upd}{\mathrm{d}}

\newcommand\e{\text{e}}
\renewcommand\i{\text{i}}
\newcommand\diff{\mathrm{d}}
\newcommand\expect[1]{\Braket{#1}}
\renewcommand\geq\geqslant
\renewcommand\leq\leqslant

\definecolor{myblue}{RGB}{94.3418,129.735, 181.708}
\definecolor{myokker}{RGB}{225.465, 156.426, 36.3651}
\definecolor{mygreen}{RGB}{143.406, 177.042, 49.8906}
\definecolor{myorange}{RGB}{236.167, 98.7203, 53.5498}
\definecolor{mypurple}{RGB}{135.293, 120.48, 179.546}

\definecolor{mycyan}{RGB}{93.1579, 158.336, 200.281}
\definecolor{myyellow}{RGB}{255, 192, 0}
\definecolor{mymagenta}{RGB}{165.792, 96.809, 157.193}

\begin{document}

\title{Continuous demixing transition of binary liquids: finite-size
  scaling from the analysis of sub-systems}

\author{Y. Pathania}
\affiliation{
  Indian Institute of Science Education and Research Mohali,
  Knowledge City, Sector 81, S. A. S. Nagar,
  Manauli-140306, India
}

\author{D. Chakraborty}
\email{chakraborty@iisermohali.ac.in}
\affiliation{
  Indian Institute of Science Education and Research Mohali,
  Knowledge City, Sector 81, S. A. S. Nagar,
  Manauli-140306, India
}

\author{F. H\"of{}ling}
\email{f.hoefling@fu-berlin.de}
\affiliation{
  Freie Universität Berlin,
  Fachbereich Mathematik und Informatik,
  Arnimallee 6,
  14195 Berlin, Germany
}
\affiliation{
  Zuse Institute Berlin,
  Takustr. 7,
  14195 Berlin, Germany
}

\date{\today}

\begin{abstract} \noindent A binary liquid near its consolute
  point exhibits critical fluctuations of the local composition;
  the diverging correlation length has always challenged simulations.
  The method of choice for the calculation of critical points in the phase diagram is a scaling analysis of
  finite-size corrections, based on a sequence of widely different system sizes.
  Here, we discuss an alternative using cubic sub-systems of one large
  simulation as facilitated by modern, massively parallel hardware.
  We exemplify the method for a symmetric binary liquid at critical composition
  and compare different routes to the critical
  temperature: (1)~fitting critical divergences
  encoded in the composition structure
  factor of the whole system, (2)~testing data collapse and scaling of
  moments of the composition fluctuations in sub-volumes, and
  (3)~applying the cumulant intersection criterion to the sub-systems.
  For the last route, two difficulties arise: sub-volumes are open systems with
  free boundary conditions, for which no precise
  estimate of the critical Binder cumulant $U_c$ is available.
  Second, the periodic boundaries of the simulation box interfere with
  the sub-volumes, which we resolve by a two-parameter
  finite-size scaling. The implied modification to the
  data analysis restores the common intersection point, and we
  estimate $U_c=0.201\pm0.001$, universal for
  cubic Ising-like systems with free boundaries.
  Confluent corrections to scaling, which arise for small sub-system sizes, are quantified at leading order and our data for the critical susceptibility are compatible with the universal correction exponent $\omega\approx 0.83$.
\end{abstract}


\maketitle

\section{Introduction}

%

The calculation of phase diagrams of fluid substances is of interest for manifold applications.
A particularly challenging task is finding the loci of critical points,
with the liquid--vapour critical point as a prominent example.
Binary liquid mixtures, in addition, exhibit phase separation and the coexistence of differently composed liquid phases, which leads to a line of critical points in the phase diagram \cite{Dietrich1997,Koefinger:JCP2006}.
In the vicinity of these critical points, the fluid is characterised by diverging length and time scales
\cite{Folk2006, Stanley1999, Beysens1982, Burstyn1980},
which puts considerable difficulties on their simulation.
The growth of the correlation length interferes with the finite size of the simulation
box, with the consequences that the observed phase transition is rounded and
shifted and critical divergences are capped.

In the past decade, the seemingly never ending growth in computing power was, among other factors, driven by a shift to massive parallelisation, making molecular simulations of unprecedented system sizes and run lengths broadly available \cite{Heinen:JCP2019, Straube:CP2020, Hofling2015, Ingebrigtsen:PRX2019, Chaudhuri:2016, Roth:APA2016, Glaser:2015}.
The use of huge systems mitigates artifacts due to a finite simulation box and allows, in principle, probing the critical divergences directly as one would do in an experiment.
Yet, an elegant and conceptually superior alternative exploits the scale invariance of the critical fluid
and turns the limitation of a finite system size into an advantage
by explicitly following the divergences of certain fluid properties as a function of the system size \cite{
Binder:1981ij,Binder:1981hj}.
The latter approach requires that the simulations at each thermodynamic state point are repeated for a wide range of system sizes, including comparably small systems.
However, on prevalent massively parallel hardware such as high-performance graphics processing units (GPUs)
it is not possible to run a set of independent simulations concurrently, and so the conventional finite-size scaling does not participate in substantial advances of computing hardware.

Both aspects, a systematic finite-size scaling and the efficient simulation of huge systems can be combined in the finite-size scaling analysis of sub-systems.
This old idea was carried out successfully for two- (2D) and three-dimensional (3D) Ising lattice models \cite{Binder:1981ij,Binder:1981hj}, but its application to 2D Lennard-Jones (LJ) fluids was limited by the computing resources available 30 years ago and suffered from the insufficient separation of the two length scales \cite{Rovere:1988, Rovere:1990, Rovere:1993}.
The approach was revived only recently to determine critical points in the phase diagrams of 3D active suspensions \cite{Trefz:JCP2017,Siebert:2018}; in these studies, additional countermeasures were needed to avoid biased sampling by sub-volumes that contained an interface.

Compared to the conventional finite-size scaling, there are two important differences: boundary conditions are not periodic, and the system geometry is inherently described by two lengths: the simulation box $L$ and the sub-system size~$\ell$.
In the limit of small sub-systems relative to the simulation volume ($\ell \ll L$), the sub-systems realise open systems in the sense of statistical mechanics; in particular, they can exchange heat and mass with a large reservoir.
Open sub-systems are receiving increasing interest in the simulation community, motivated by studies of small-system thermodynamics \cite{Schnell:CPL2011,Kjelstrup:ANSNN2014} but also liquid--vapour interfaces \cite{SSFslab:JCP2020} and by the finite-size scaling of Kirkwood--Buff integrals to calculate chemical potentials \cite{CortesHuerto2016,Heidari2018}.
On the other hand, methodological advances in adaptive resolution techniques permit the direct simulation of an open system coupled to one \cite{Fritsch:PRL2012,DelleSite:PR2017,DelleSite:2019} or many \cite{EbrahimiViand:JCP2020} reservoirs acting as a thermodynamic mean field.
The free boundary conditions, as realised by sufficiently small, open sub-systems, modify the fluctuations of density and composition \cite{SSFslab:JCP2020}.

In the context of critical phenomena, the critical amplitudes are known to be different for free and periodic boundary conditions \cite{Privman1984,Brezin1985}, which applies also to the critical Binder cumulant $U_c$ used to locate the critical temperature
\cite{Binder:1981ij,Binder:1981hj}.
Specifically, the value \cite{Bloete1995,Bloete1999} $U_c = 0.6236\pm 0.0002$
is universal for cubic Ising-like systems with periodic boundaries
and thus relevant for conventional finite-size scaling based on a
sequence of simulations \cite{Kim2003,Das2006}.
For free boundary conditions, however, only a rough estimate $U_c\approx 0.21$
exists \cite{Binder:1981ij}, and this lack of knowledge of $U_c$
constitutes an obstacle to the
straightforward application of sub-volume finite-size scaling.
Another obstacle lies in the competition of the two lengths $\ell$ and $L$, i.e., the size of the sub-system and that of the simulation box, which renders the effective boundary conditions non-ideal, i.e., neither free nor periodic.

In this work, we revisit the sub-system scaling analysis in the vicinity of a critical point and compare different approaches to estimate the critical temperature.
We introduce a two-parameter scaling ansatz that accounts explicitly for the two lengths $\ell$ and $L$,
thereby suggesting a finite-size protocol suitable for finite simulation boxes.
As a by-product, we improve the estimate of $U_c$ for free boundaries by an order of magnitude.
We demonstrate the method for a symmetric binary liquid of Lennard-Jones particles, which has pair interactions that are symmetric with respect to the two molecular species A and B. Thus, the critical composition is a 1:1 mixture that exhibits phase separation into symmetric A-rich and B-rich phases, occurring in close analogy to the spontaneous symmetry breaking of the Ising model.
This model system (and slight variants thereof) was studied extensively near consolute points using combinations of semi-grand canonical Monte Carlo (SGMC) methods and molecular dynamics (MD) simulations \cite{Das2003,Das2006,Das2006a,Ahmad:2012gm,Roy:2011jl,Das:2012il,Roy:JCP2016}.
In particular, cuts of the phase diagram in the temperature--composition and in the temperature--density planes
are available, and the critical fluctuations of the local concentration were shown to scale as expected for the 3D-Ising universality class.
Moreover, the coarsening kinetics and the critical singularities of the transport coefficients were characterised in great detail, corroborating theoretical expectations and being in agreement with the available experimental evidences.

After giving technical details on the model liquid and the simulations in \cref{sec:model_system},
we analyse in \cref{sec:corrlength_susceptibility} the critical behaviour of the structure factor $S_\mathrm{cc}(k)$ of local concentration fluctuations obtained from the whole, large simulation box for a range of wavenumbers~$k$.
This is then contrasted in \cref{sec:sub-system_scaling} by a scaling analysis of the concentration susceptibility $\chi_\ell$ calculated from sub-volumes of different edge lengths~$\ell$.
Both approaches yield already estimates of the critical temperature $T_c$.
Binder's cumulant method based on sub-systems is carried out in \cref{sec:binder_cumulant} to improve these values
and we show the necessity for the two-parameter scaling ansatz.

\section{Model System and Simulation Details}
\label{sec:model_system}

The model of a symmetric binary mixture considered here \cite{Das2006, Roy:2011jl} employs pair interactions given by the truncated and force-shifted Lennard--Jones potential
\begin{equation}
  u_{\alpha \beta}(r) =
    u_{\text{LJ};\alpha \beta}(r)
    -u_{\text{LJ};\alpha \beta}(r_c)
    -(r-r_c) \, u'_{\text{LJ};\alpha \beta}(r_c)
\end{equation}
for $r \leq r_c$ with
$
 u_{\text{LJ};\alpha \beta}(r)=4\epsilon_{\alpha\beta} \mleft[
   (r/\sigma_{\alpha\beta})^{-12} -(r/\sigma_{\alpha\beta})^{-6} \mright]
$
as usual for particle species' $\alpha, \beta  \in \{ A , B \}$.
All particles have the same diameter,
$\sigma_{\alpha \beta}=:\sigma$, and mass $m$, but different
interaction strengths, $\epsilon_{AA} = \epsilon_{BB} =2\epsilon_{AB} =: \epsilon$.
The number density $\rho$ of the fluid is fixed at
$\rho \sigma^{-3} = 1$ in order to reduce interferences with the
liquid--vapour critical point.  At critical composition ($N_A = N_B$),
this binary fluid exhibits a continuous demixing transition at the
critical temperature \cite{Das2006, Roy:2011jl}
$T_c \approx 1.421 \epsilon/\kb$.
Dimensionless quantities are indicated by an asterisk and are formed with
$\epsilon, \sigma, \tau:=\sqrt{m \sigma^2/\epsilon}$
as units of energy, length, and time; for example,
$\rho^* := \rho\sigma^3$ and $T^*:= \kb T/\epsilon$.

A typical MD workflow consists of thermalisation and equilibration
runs, followed by one or several production runs.  For temperature
control, we used a Nos\'e--Hoover thermostat as described in
Ref.~\citenum{Melchionna1993} for the initial equilibration in the
canonical ensemble ($N_A$, $N_B$, $V$, $T$ fixed).  Then, the
thermostat was switched off and the system evolved microcanonically,
i.e., at fixed total energy, with a time step of $\delta t =0.001 \tau$.
The subsequent production run was
performed microcanonically as well and particle configurations and 
observables such as Fourier modes of the density field were recorded every $10^5$ integration steps. A single
simulation run comprised of an initial thermalisation for
$20{\,}000 \tau$, followed by microcanonical equilibration and production runs (at fixed total energy $E$) over $50{\,}000 \tau$ and $30{\,}000 \tau$, respectively.
Further, for each temperature, the data were averaged
over two independent runs
and the error bars shown in the figures represent the standard errors of these means.
The microcanonical ensemble was used because we have also calculated transport coefficients and time correlation functions (not discussed here) from the same set of simulations; the intermediate equilibration step is needed for proper temperature control in the NVE ensemble.
For the cubic simulation box, periodic
boundary conditions and an edge length of $L \approx 44.4 \sigma$ were
chosen, corresponding to a total of $87{\,}808$ particles.
The home-grown code takes advantage of
the massively parallel architecture of high-end graphics processing
units (GPU) \cite{Chakraborty2011c,Chakraborty2011b,Rings2012a},
and simulations were run on a GPU of type Tesla K20Xm (Nvidia Corp.).

Simulations for \cref{fig:susc_corr_Tc} were performed at the anticipated critical temperature $T^*=1.421$ with $N=108{\,}000$ particles ($L\approx 47.6\sigma$) using the software \emph{HAL's MD package} \cite{Colberg2011,HALMD}.
Particle configurations were stored every $10\tau$ in the structured, compressed, and portable H5MD format \cite{H5MD:2014}.
Statistical data were averaged over three independent simulation cycles.
Each cycle started with an initial thermalisation at the target temperature over $1.5\times 10^5 \tau$ and a subsequent microcanonical equilibration of the same duration.
Then the velocities of the final configuration were rescaled to impose a total (or: internal) energy of $E_\mathrm{tot} = -0.616\epsilon$, which corresponds to $T_c$.
After further equilibration over $10^5\tau$, a data production run of the same length followed.
This procedure allowed us to control the temperature of the microcanonical simulation with an accuracy of $\Delta T = 0.0003/\kb$.
Each cycle required about 47\,h on a single GPU of type Tesla V100-PCIe.

\section{Correlation length and concentration susceptibility}
\label{sec:corrlength_susceptibility}

\begin{figure}
\centering 
\includegraphics[width=\linewidth]{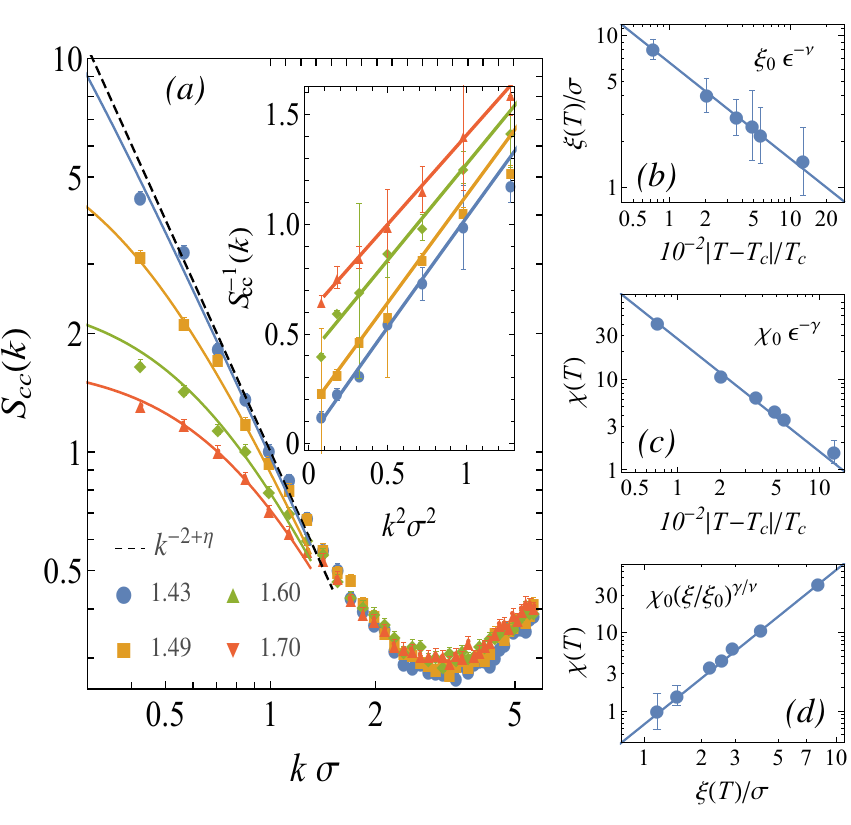}
\caption{(a)~Static structure factor $S_\mathrm{cc}(k)$ of the binary fluid at critical composition studied
  for selected temperatures approaching the consolute point from above. The reduced temperatures $T^*=\kb T/\epsilon$ are given in the legend. The data are shown on double-logarithmic scales, the inset displays a rectification of the data according to
  \cref{eq:oz_relation}. 
  Solid lines are fits to \cref{eq:oz_relation,eq:oz_fisher_relation}, respectively (see text), and the dashed line indicates the critical divergence, $S_\mathrm{cc}(k\to 0) \sim k^{-2+\eta}$ at $T=T_c$.
  ~~(b--d)~Critical divergences of the correlation length $\xi(T)$ and the concentration susceptibility $\chi(T)$,
  the data points along with their uncertainties were obtained from linear regressions to $S_\mathrm{cc}(k)^{-1}$ vs. $k^2$ as shown in the inset of panel~(a).
  Solid lines are power-law fits using the known critical exponents $\nu$ and $\gamma$ (see main text) to estimate the critical temperature $T_c$ and the non-universal amplitudes $\xi_0$ and $\chi_0$, respectively.
  }
\label{fig:scck}
\end{figure}

The structural properties of the system arise from two fluctuating
fields: the density field
$\delta \rho(\vec{r})=\delta \rho_A(\vec{r})+\delta \rho_B(\vec{r})$
and the concentration field
$\delta c(\vec{r})=x_B \delta \rho_A(\vec{r})-x_A \delta
\rho_B(\vec{r})$,
where $\delta \rho_\alpha(\vec{r})$ denotes the fluctuations of the
microscopic, partial density of species $\alpha$,
$x_\alpha = N_\alpha / N$ is the mole fraction of the species, and
$N = N_A + N_B$.  The relevant order parameter is
$\varphi=\langle |x_A-x_c| \rangle $, where the critical composition
is $x_c=1/2$ due to the symmetry of the mixture under investigation.

Critical fluctuations of the composition are quantified by the
generalised static structure factor,
$S_\mathrm{cc}(k)=N\mleft\langle |\delta c(\vec{k})|^2 \mright\rangle$ with
$\delta c(\vec{k}):=\int_V \e^{\i\vec{k}\cdot \vec{r}} \, \delta
c(\vec{r}) \, \diff^3 r$,
which far away from the critical point is of the Ornstein--Zernike form for small wavenumber~$k$,
\begin{equation}
  \label{eq:oz_relation}
  S_\mathrm{cc}(k) \simeq \rho \kb T \chi \mleft[1+ (k\xi)^2 \mright]^{-1} \,; \quad
  k\sigma \ll 1 \,.
\end{equation}
This asymptotic expression serves as definition for the (second-moment) correlation length $\xi(T)$
and the concentration susceptibility $\chi(T)$.
Upon approaching the consolute point, the two quantities exhibit the familiar algebraic divergences \cite{Stanley1971}
\begin{equation}
  \xi(T) \simeq \xi_0 \varepsilon^{-\nu} \quad \text{and} \quad
  \chi(T) \simeq \chi_0 \varepsilon^{-\gamma} \,; \quad T \downarrow T_c,
  \label{eq:xi_chi_divergences}
\end{equation}
as functions of the reduced temperature $\varepsilon:=|T-T_c|/T_c$.
These critical laws are governed by the short-ranged Ising universality
universality class for three dimensions, which fixes the exponents to their precisely known values, $\nu\approx 0.630$ and $\gamma \approx 1.237$, respectively \cite{Pelissetto2002};
the non-universal amplitudes $\xi_0$ and $\chi_0$ are system-specific.
\footnote{Note that depending on whether $T_c$ is approached from above or below, different critical amplitudes apply, e.g., $\xi_0^+$ and $\xi_0^-$.
Our analysis of $\xi(T)$ and $\chi(T)$ is restricted to $T > T_c$, why we refrain from indicating the $+$ sign at the amplitudes to keep the notation simple.}
Moreover, the static structure factor assumes the scaling form
\begin{equation}
  S_\mathrm{cc}(k) = k^{-2+\eta} s(k\xi) \,;  \quad k \sigma \ll 1, \: \xi/\sigma \gg 1,
\end{equation}
where the anomalous dimension $\eta =2-\gamma/\nu \approx 0.036$ characterises the spatial decay of
the OP correlations and $s(x)$ is a universal scaling function, which
interpolates from the critical power-law, $s(x \gg 1) = \textit{const}$,
to the convergence of $S_\mathrm{cc}(k)$ at small wavenumber, $s(x \to 0) = x^{2-\eta}$;
the latter encodes the scaling of the susceptibility.
The anomalous scaling of $S_\mathrm{cc}(k)$ requires a modification of \cref{eq:oz_relation} to
the form,\cite{Fisher1964} 
\begin{equation}
 \label{eq:oz_fisher_relation}
 S_\mathrm{cc}(k) \simeq \rho \kb T \chi \mleft[1+ (k\xi)^2 \mright]^{-1+\eta/2}.
\end{equation}

Our simulation data for $S_\mathrm{cc}(k)$ are well described by relations
\cref{eq:oz_relation,eq:oz_fisher_relation}, respectively, for
temperatures ranging from $T^* = 1.70$ down to $1.43$, see
\cref{fig:scck}(a).
Linear regression of
$S_\mathrm{cc}(k)^{-1}$ as function of $(k\sigma)^2$ for
$k\sigma \lesssim 1$ yields values for $\xi$
and $\chi$ (see inset);
\cref{eq:oz_fisher_relation} was used for $T^* \leq 1.60$.
The data nicely follow the expected critical singularities
over more than one decade in magnitude and for $\varepsilon \lesssim 0.1$ [see
\cref{fig:scck}(b)--(d)].
Fitting both power laws to the respective data
set provides us with first estimates of the critical temperature $T_c$
as well as the amplitudes $\xi_0$ and $\chi_0$.  Using the correlation
length data, we obtained $T_c^*=1.421 \pm 0.001$ and
$\xi_0=(0.365 \pm 0.009)\sigma$, while the data for the
concentration susceptibility yielded $T_c^*=1.421 \pm 0.001$ and
$\chi_0^*=0.092 \pm 0.011$.  The results for $T_c$ and $\xi_0$ are in
excellent agreement with known results \cite{Das2006, Roy:2011jl}.  
Merely our value for $\chi_0^*$ is somewhat larger than the previously reported value
\cite{Das2006} of $\chi_0^*=0.076 \pm 0.006$.

\section{Scaling analysis of sub-system fluctuations}
\label{sec:sub-system_scaling}

\subsection{Concentration susceptibility}
\label{subsec:chi_sub-system_scaling}

In a typical semi-grand canonical Monte Carlo simulation, the
fluctuations in the concentration are sampled by switching the
particle identities at a predefined rate while keeping the total
particle number fixed.  Then by linear response theory, the
susceptibility $\chi$ in the mixed phase is proportional to the
variance of the fluctuating mole fraction $x_A = N_A / N$ of, e.g.,
species~$A$:
\begin{equation}
  \label{eq:conc_susceptibility}
  \rho \kb T \chi = N \expect{(x_A - x_c)^2}_\mathrm{gc} \,,
  \qquad T > T_c \,,
\end{equation}
provided that the symmetric mixture is at critical composition,
$\expect{x_A}_\mathrm{gc} = \expect{x_B}_\mathrm{gc} = x_c = 1/2$; 
in practice, the averages are taken in the semi-grand canonical ensemble \cite{Das2003,Das2006,Roy:2011jl,Roy:JCP2016}.
The standard procedure for the finite-size scaling analysis is then based
on a sequence of simulations for different boxes of linear extent $L$ to
determine $\chi(T;L)$ for each value of~$L$ and to study the behaviour as $L\to \infty$.

Above, we extrapolated the composition structure factor 
by virtue of its specific wavenumber dependence, \cref{eq:oz_fisher_relation},
to access the susceptibility,
$S_\mathrm{cc}(k \to 0) \simeq \rho \kb T \chi$.
Such a systematic change of the wavenumber has a close analogy to the described finite-size scaling analysis.
A finite system of size $L$ does not support fluctuations on length scales larger than~$L$
and, likewise, the correlations contained in $S_\mathrm{cc}(k)$ for given $k$ stem essentially
from fluctuations on length scales $\lesssim 2\pi / k$.
Thus, the value $S_\mathrm{cc}(k)/\rho \kb T$ may be interpreted
as the effective susceptibility of a finite system of size $2\pi/k$.
Observing that a single simulation yields data for different wavenumbers
suggests to perform the conventional scaling analysis but for a sequence of finite sub-systems
\cite{Binder:1981ij,Binder:1981hj}.

The idea is easily linked to \cref{eq:conc_susceptibility} by noting that sub-volumes of
a large system represent open systems coupled to a (finite) particle reservoir.
Within each sub-volume, particle number and concentration exhibit similar
fluctuations as in the grand canonical ensemble, although they are still locally
conserved quantities. \cite{Heidari2018,SSFslab:JCP2020}.
Specifically, we partitioned the simulation box of linear extent $L$ into $m^3$ cubic
sub-systems of edge length $\ell = L / m$, and recorded the fluctuating
particle numbers $N_A^{(\ell)}$ and $N_B^{(\ell)}$ for each sub-system;
the total number of particles is denoted by $N_\ell := N_A^{(\ell)} + N_B^{(\ell)}$.
The susceptibility $\chi_\ell$ is then calculated from
\cref{eq:conc_susceptibility} by substituting $x_A = N_A^{(\ell)} / N_\ell$ 
for the instantaneous concentration and replacing $N$ by the average particle number $\expect{N_\ell} = N / m^3$ in
each sub-system.

\begin{figure*}
\centering
\includegraphics[width=\linewidth]{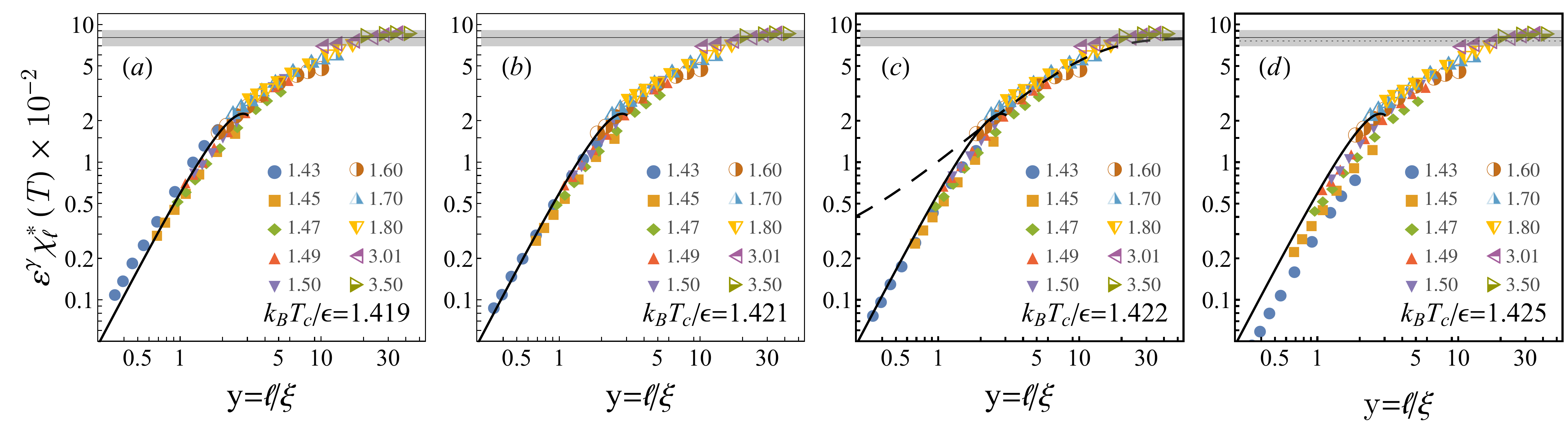}
\caption{Scaling plots of the reduced susceptibility $\chi^*(T) := \chi(T) \,\epsilon/\sigma^3$
  according to \cref{eq:susceptibility_scaling} with the correlation length $\xi=\xi(T)$ obtained from the static structure factor as in \cref{fig:scck}.
  The data collapse onto the scaling function $\chi_0 Z(y)$ is tested for four different choices of the critical temperature $T_c$ as written in each panel, which enters the reduced temperature $\varepsilon := (T-T_c)/T_c$.
  Black solid lines represent the critical asymptote as $y \to 0$, amended by the leading universal correction
  [\cref{eq:scaling_function_susceptibility_small_y}],
  which employs the amplitudes $z_0, z_1$ from the fit to $f_2$ in \cref{fig:s2_s4_combined}.
  The black dashed line in panel (c) is a fit to the large-$y$ approximation [\cref{eq:scaling_function_susceptibility_large_y}] of $\chi_0 Z(y)$
  and indicates the approach to the Gaussian fixed point, $y \to \infty$.
  The amplitude $\chi_0$ (horizontal black line) estimated from this fit and its uncertainty (gray shaded region) are indicated.
  }
  \label{fig:susceptibility_collapse}
\end{figure*}

The standard finite-size scaling of the susceptibility  \cite{Stanley1971,Das2006}
suggests that $\chi_\ell$ for $T > T_c$ asymptotically obeys the scaling ansatz
\begin{equation}
  \label{eq:susceptibility_scaling}
  \chi_\ell(T) \simeq \chi_0 \, \varepsilon^{-\gamma} \, Z(\ell/\xi)\,, \quad
  \ell, \xi \gg \sigma \,,
\end{equation}
where $\varepsilon = \varepsilon(T)$ as before and $\xi = \xi(T)$ serves as
a short-hand for the power law divergence given in \cref{eq:xi_chi_divergences}.
The function $Z(\cdot)$
is the appropriate scaling function that interpolates between the thermodynamic
critical divergence ($\ell \to \infty$, $\xi$ fixed) and the finite-size scaling
at criticality ($\xi \to \infty$, $\ell$ fixed).
\footnote{To simplify the discussion here, we have tacitly assumed an
arbitrarily large simulation box that supports the limit $\ell \to \infty$. See
\cref{sec:finite-size-corrections} for an alternative.}
In the first limit, the scaling variable $y :=\ell/\xi \to \infty$ and
the scaling function approaches unity exponentially fast \cite{Das2006},
and a satisfactory description of the data for $y \gtrsim 3$ follows the form
\begin{equation}
  \label{eq:scaling_function_susceptibility_large_y}
  Z(y \to \infty) \simeq \left(1 - z_\infty \e^{-a_1 y} \right)\,,
\end{equation}
including the leading correction to scaling with $z_\infty$ and $a_1$ being 
system-specific parameters (see \cref{fig:susceptibility_collapse,fig:s2_s4_combined}).

For the opposite limit $y\to 0$, we
recall that for a finite system, all physical observables are analytic
in the temperature, also at $T_c$. For the susceptibility, it
implies $\chi_\ell(T) = \chi_\ell(T_c) + O(\varepsilon)$ as
$\varepsilon \to 0$ such that by comparing with
\cref{eq:susceptibility_scaling} the scaling function behaves as
\begin{equation}
  \label{eq:scaling_function_susceptibility_small_y}
  Z(y \to 0) \simeq z_0 y^{\gamma/\nu} \left( 1- z_1 y^{1/\nu} \right) ,
\end{equation}
which introduces the universal amplitudes~$z_0$ and $z_1$. 
Combining with $\xi \simeq \xi_0 \varepsilon^{-\nu}$ and specialising
to $T=T_c$, one recovers the finite-size scaling form well-known for
periodic boundaries,
\begin{equation}
  \label{eq:susceptibility_critical_scaling}
  \chi_{\ell}(T_c) \simeq z_0 \chi_0 \, \left(\frac{\ell}{\xi_0}\right)^{\gamma/\nu} \,,
  \quad \ell \gg \sigma.
\end{equation}
More generally, one finds for the critical region near $T_c$ a linear dependence
on $T$:
\begin{equation}
  \chi_{\ell}(T) \simeq \tilde \chi_0 \, \ell^{\gamma/\nu}
    \mleft[ 1- \tilde z_1 \ell^{1/\nu}\, (T-T_c) \mright]
\end{equation}
with coefficients $\tilde\chi_0 :=z_0 \chi_0 \xi_0^{-\gamma/\nu}$ and
$\tilde z_1 := z_1 \xi_0^{-1/\nu} T_c^{-1}$
for the regime $\sigma \ll \ell \ll \xi(T)$.
Confluent corrections to scaling are relevant for small $\ell$; their discussion is deferred to \cref{sec:confluent}.

The scaling form \cref{eq:susceptibility_scaling} can be used to
determine the critical temperature $T_c$, the amplitude $\chi_0$, and
the scaling function $Z(y)$.  Plotting the data for
$\varepsilon^{\gamma} \chi_\ell(T)$ for a range of temperatures and
sub-system sizes as a function of $y=\ell/\xi$ should yield data
collapse onto the function $\chi_0 Z(y)$.
\Cref{fig:susceptibility_collapse} shows a sequence of such plots for
different estimates of $T_c$, which are informed by our previous
result from \cref{sec:corrlength_susceptibility}.  The quality of the
data collapse for the different choices of $T_c$ gives us a fairly
good estimate of the critical temperature.  Comparing panels (b) and
(c) of \cref{fig:susceptibility_collapse}, in particular close to
$y=\ell/\xi \approx 0.8$, suggests that the critical temperature is
close to $T_c^*=1.422$ [\cref{fig:susceptibility_collapse}(c)].

A more quantitative estimate is obtained by using the form of $Z(y)$ as
$y \to 0$. However, this requires knowledge of the parameters $z_0$ and
$z_1$, which can independently be obtained from the sub-system scaling
of the second moment of the order parameter $\phi:=x_A - x_c$
as discussed in \cref{subsec:finite_size_scaling_order_parameter}.
Specifically, the quantity $a^2 \ell^{2\beta/\nu} \expect{\phi^2}_\ell$
in the limit of $y=\ell/\xi\to 0$ has the scaling form $z_0(1-z_1
y^{1/\nu})$; for the definition of $a$ see \cref{eq:a} below.
Fitting this functional form to the data for $\expect{\phi^2}_\ell$,
scaled by $a^2 \ell^{2\beta/\nu}$, within the range
$0.34 \leq y \lesssim 0.78$ yields $z_0=0.095 \pm 0.004$ and
$z_1=0.16 \pm 0.05$ (\cref{fig:s2_s4_combined}). 
Equipped with these values, we plot the scaling function $Z(y)$
for $y \to 0$ (solid black lines in \cref{fig:susceptibility_collapse}), 
which guides the assessment of the quality of the data collapse.
The definition of $a$ contains the critical temperature $T_c$, 
why our estimate $T^{*}_c=1.421$
from the susceptibility data given in
\cref{sec:corrlength_susceptibility} was used to fix $a$. It is
worth noting that from the data collapse of $\chi_\ell$, the value
of $T_c$ obtained here agrees very well with that obtained from the
critical divergences of the correlation length and the susceptibility,
given the measurement uncertainty of the values of $\xi$ entering 
the scaling plots in \cref{fig:susceptibility_collapse}.

For this value of $T_c$, the rescaled data represent the function
$\chi_0 Z(y)$ (\cref{fig:susceptibility_collapse}c).
For values of $y=\ell/\xi \gtrsim 20$, the data indeed
converge and, in principle, one could read off $\chi_0$ directly.
However, a more robust approach should allow for corrections to
scaling, \cref{eq:scaling_function_susceptibility_large_y}, which
suggests to fit this form to the data with $\chi_0, z_\infty$
and $a_1$ as parameters. Even though the range of values for large $y$
is severely limited, we performed an asymptotic fit as $y \to \infty$ to
estimate the amplitude $\chi_0$, yielding
$\chi_0 = 0.080 \pm 0.002$ and thereby improving our earlier estimate of
$\chi_0=0.092 \pm 0.011$. The value of $a_1$ from the best fit comes
out to be $a_1=0.11 \pm 0.02$ and that of $z_\infty=1.0 \pm 0.3$.

\subsection{Order parameter distribution}
\label{subsec:finite_size_scaling_order_parameter}

The concentration susceptibility $\chi$ is essentially the variance of
the fluctuating order parameter $\phi := x_A - x_c$, see
\cref{eq:conc_susceptibility}, and thus merely one characteristic of
the statistical distribution of $\phi$.  A more general description is
in terms of the probability density $P_\ell(\phi;T)$ of $\phi$ for
sub-system size $\ell$ and temperature $T$,
which also admits a finite-size scaling hypothesis
\cite{Binder:1981ij,Binder:1981hj}:
\begin{equation}
  \label{eq:order_parameter_dist}
  P_\ell(\phi;T) \simeq
    a \ell^{\beta/\nu}
    \mathcal{P}\mleft(a \ell^{\beta/\nu} \phi, \ell/\xi(T) \mright) \,,
  \qquad \ell, \xi(T) \gg \sigma,
\end{equation}
with a universal scaling function $\mathcal{P}(\hat y,y)$, where
$\hat y:=a \ell^{\beta/\nu} \phi$ and $y:=\ell/\xi(T)$, and a scale
factor $a$ such that 
\begin{equation}
  a^{-2} = \kb T_c \chi_0 \xi_0^{-\gamma/\nu};
  \label{eq:a}
\end{equation}
as usual, the critical exponent \cite{Pelissetto2002} $\beta\approx 0.326$ describes the scaling of the order parameter as $T\uparrow T_c$.

By normalisation of $P_\ell$, it holds
$\int_{-\infty}^\infty \mathcal{P}(\hat y, y) \,\diff\hat y = 1$ for
all $y$.  At critical composition of the mixture,
$\expect{x_A}_\ell = x_c$, the first moment of $P_\ell$ vanishes due
to symmetry, $\expect{\phi}_\ell = 0$.  The second moment encodes the
scaling of the susceptibility,
\begin{align}
  \chi_\ell(T) &= \frac{\ell^d}{\kb T} \int \! \diff\phi \, \phi^2 \, P_\ell(\phi; T) \\
               &\simeq \frac{\ell^{d-2\beta/\nu}}{a^2 \kb T} \, \int \!\diff \hat y \, \hat y^2 \,\mathcal{P}(\hat y, \ell/\xi) \,,
\end{align}
by virtue of the definition of $\chi_\ell$ analogous to
\cref{eq:conc_susceptibility}, with $\expect{N_\ell} = \rho \ell^d$,
and using \cref{eq:order_parameter_dist} in the second step.  For
every value of $y = \ell/\xi$, the $\hat y$-integral yields a constant
$f_2(y)$. Invoking the hyper-scaling relations \cite{Pelissetto2002}
$\gamma/\nu = 2 - \eta = d - 2\beta/\nu$, we recover
\cref{eq:susceptibility_scaling}:
\begin{align}
  \chi_\ell(T) &\simeq \frac{\ell^{\gamma/\nu}}{a^2 \kb T} f_2(\ell/\xi) \\
    &\simeq \chi_0 \frac{\varepsilon^{-\gamma}}{1+\varepsilon}\, Z(\ell/\xi) \,,
    &  \ell, \xi \gg \sigma
\end{align}
upon identifying the scaling function as
$Z(y) = y^{\gamma/\nu} f_2(y)$, substituting
$\xi=\xi_0 \varepsilon^{-\nu}$ and $\kb T = \kb T_c(1+\varepsilon)$,
and due to our choice of~$a$.

\begin{figure}
 \includegraphics[width=\figwidth]{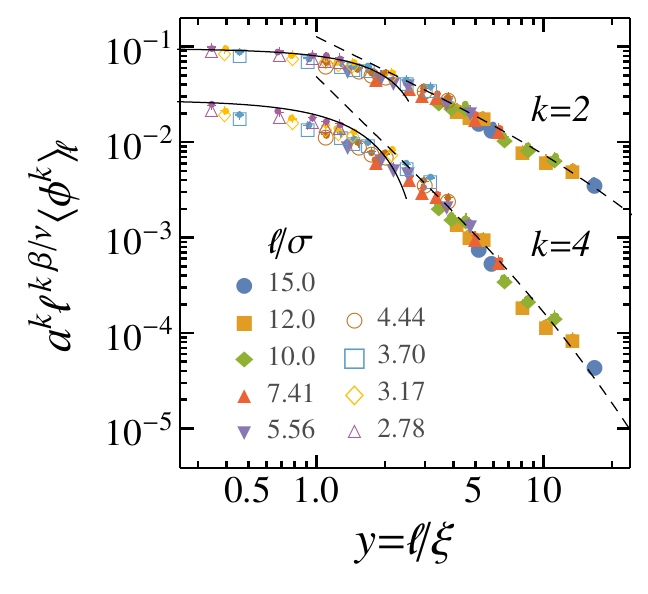}
 \caption{Scaling of the 2\textsuperscript{nd} and 4\textsuperscript{th} moments
   of the concentration fluctuations $\phi = x_A - x_c$ within the sub-volumes.
   The data collapse onto the scaling functions $f_k(y)$ is tested by plotting 
   $a^k \ell^{k\beta/\nu}\langle \phi^k \rangle_\ell$ against $y=\ell/\xi(T)$ for $k=2$ and $k=4$, respectively,
   with the critical exponents $\beta$ and $\nu$ given by the Ising universality class (see text),
   the correlation length $\xi=\xi(T)$ obtained from $S_\mathrm{cc}(k)$ [\cref{fig:scck}],
   and $a$ being a constant factor [\cref{eq:a}].
   The different symbols refer to data from the sub-volume sizes $\ell/\sigma$ specified in the legend.
   Lines show the small- and large-$y$ asymptotes of the scaling functions $f_2(y)=y^{-\gamma/\nu}Z(y)$ and, within the Gaussian approximation, of $f_4(y) \simeq 3 f_2(y)^2$; the latter becomes exact as $y\to \infty$.
   \Cref{eq:scaling_function_susceptibility_large_y} was used for the large-$y$ behaviour of $Z(y)$ (solid lines),
   whereas the critical asymptote $Z(y \to 0)$ is given in \cref{eq:scaling_function_susceptibility_small_y} (broken lines);
   see also the lines in \cref{fig:susceptibility_collapse}(c).
}
\label{fig:s2_s4_combined}
\end{figure}

Similarly, one readily obtains for the $k^\mathrm{th}$ moment of the order parameter the scaling form
\begin{align}
  \label{eq:op_moments}
  \bigl\langle \phi^k \bigr\rangle_\ell
  &:= \int \! \diff \phi \, \phi^k \, P_\ell(\phi; T) \\
  &\simeq a^{-k} \ell^{-k\beta/\nu}
      \int_{-\infty}^{\infty} \upd \hat y ~\hat y^k~\mathcal{P}(\hat y, \ell/\xi). \\
  &=: a^{-k} \ell^{-k \beta/\nu} f_k(\ell/\xi),
  \label{eq:op_scaling}
\end{align}
which defines universal scaling functions $f_k$ for $k=1,2,\dots$ If
$P_\ell(\phi)$ is symmetric,
$\bigl\langle \phi^k \bigr\rangle_\ell = 0$ for $k$ odd.  In
\cref{fig:s2_s4_combined}, we test this scaling prediction for $k=2$
and $4$ on the sub-system analysis of the present binary fluid:
plotting $a^k \ell^{k \beta/\nu}\langle\phi^k\rangle_\ell$ against
$\ell/\xi$ the data collapse nicely onto the functions $f_k$ over
the full range $0.05 \lesssim \ell/\xi \lesssim 20$.

In the regime $y \gg 1$, i.e., for large, near-critical sub-systems,
the distribution $P_\ell(\phi)$ is Gaussian, which is inherited to the
scaling function $\mathcal{P}(\cdot,y)$ being Gaussian in its first
argument with zero mean and variance $f_2(y)$.  In this case, the
moments and thus the functions $f_k$ are related to each other since
all cumulants except the first two vanish.  Specifically,
$\expect{\phi^4}_\ell \simeq 3\expect{\phi^2}_\ell^2$ for
$\ell \gg \xi \gg \sigma$, which implies
\begin{equation}
  f_4(y \to \infty) \simeq 3 f_2(y)^2 = 3 y^{-2\gamma/\nu} Z(y)^2 \,.
\end{equation}
For comparison, the functions $f_2$ and $f_4$ as predicted from the
small- and large-$y$ approximations to $Z(y)$ (\cref{fig:susceptibility_collapse}) have been
included in \cref{fig:s2_s4_combined}. Both functions describe the 2\textsuperscript{nd}
and 4\textsuperscript{th} moments very well for the whole range of $y$-values available.
Merely for $y \lesssim 1$, the data for $k=4$ deviate visibly from
$f_4(y)$, indicating a non-Gaussian distribution as expected close to
criticality.

\subsection{Confluent corrections to scaling}
\label{sec:confluent}

In the preceding sections, we discussed the asymptotic behaviour for large sub-system sizes $\ell \gg \sigma$.
However, large sub-system sizes are challenging to reach in simulations and a question of practical importance is how quickly does an observable such as $\chi_\ell(T)$ approach its leading power-law asymptote as $\ell$ is increased?
Renormalisation group (RG) theory explains how these so-called confluent corrections to scaling emerge from irrelevant scaling variables \cite{Cardy:Scaling,Pelissetto2002}.
These variables encode microscopic details of the system that fade out upon coarse-graining by the RG flow;
yet, the confluent corrections are associated with universal critical exponents.
This type of corrections was analysed in simulation studies of, e.g., the 3D Ising model \cite{Hasenbusch:PRB2010,Xu:JPCS2018}, the 3D Heisenberg model \cite{Holm:PRB1993}, the statistics of percolation clusters \cite{Ziff:2011,Percolation_EPL:2008},
and critical transport on such structures \cite{Lorentz_PRL:2006,Percolation_EPL:2008}.

For the susceptibility, we include the leading confluent correction in our discussion by extending the scaling ansatz \cref{eq:susceptibility_scaling} by an irrelevant scaling variable $u$
\begin{equation}
  \label{eq:susceptibility_scaling_corrections}
  \chi_\ell(T) \simeq \chi_0 \, \varepsilon^{-\gamma} \, \mathcal{Z}(\ell/\xi,u \ell^{-\omega})\,, \quad
  \ell, \xi \gg \sigma \,,
\end{equation}
with the 3D-Ising correction exponent \cite{Pelissetto2002,ElShowk:JSP2014,Hasenbusch:PRB2010,Xu:JPCS2018} $\omega \approx 0.83$;
here, $y=\ell/\xi$ means $y=(\ell/\xi_0) \varepsilon^{\nu}$.
For finite systems, the function $\chi_\ell$ is analytic in $T$, but also in $u$;
in particular, the scaling function $\mathcal{Z}$ is analytic in its second argument and obeys $\mathcal{Z}(y, 0) = Z(y)$.
As a consequence, the amplitudes $z_0, z_1$, and $z_\infty$ in \cref{eq:scaling_function_susceptibility_large_y,eq:scaling_function_susceptibility_small_y} depend on $u \ell^{-\omega}$ and can be expanded in this parameter for large $\ell$ (keeping $u$ fixed).
At leading order, this amounts to replacing $z_i$ by $z_i (1+ \tilde c_i \ell^{-\omega})$ for $i = 0, 1, \infty$; the value of $u$ has been absorbed in the amplitudes $\tilde c_i$.
The procedure turns \cref{eq:scaling_function_susceptibility_small_y} into
\begin{multline}
  \label{eq:susceptibility_corrections}
  \mathcal{Z}(y, u\ell^{-\omega})
   \simeq z_0 y^{\gamma/\nu} \Bigl( 1- z_1 y^{1/\nu} + \tilde c_0\ell^{-\omega} \\ - z_1 \tilde c_0 y^{1/\nu} \ell^{-\omega} \Bigr)
\end{multline}
for $y\to 0$ and $\ell \gg \sigma$.
Clearly, there are two types of leading corrections to the critical divergence of $\chi_\ell$:
one type scales as $y^{1/\nu}$ and is removed by fine-tuning of the temperature to its critical value.
Second, the confluent corrections, $\sim \ell^{-\omega}$, which disappear for sufficiently large sub-system size~$\ell$.
Analytic corrections $O\bigl(\ell^{-1}\bigr)$, e.g., due to the non-linear mixing of scaling fields, do not contribute at leading order.
Corrections due to a finite simulation box will be discussed in \cref{sec:finite-size-corrections}.

\begin{figure}
 \includegraphics[width=\figwidth]{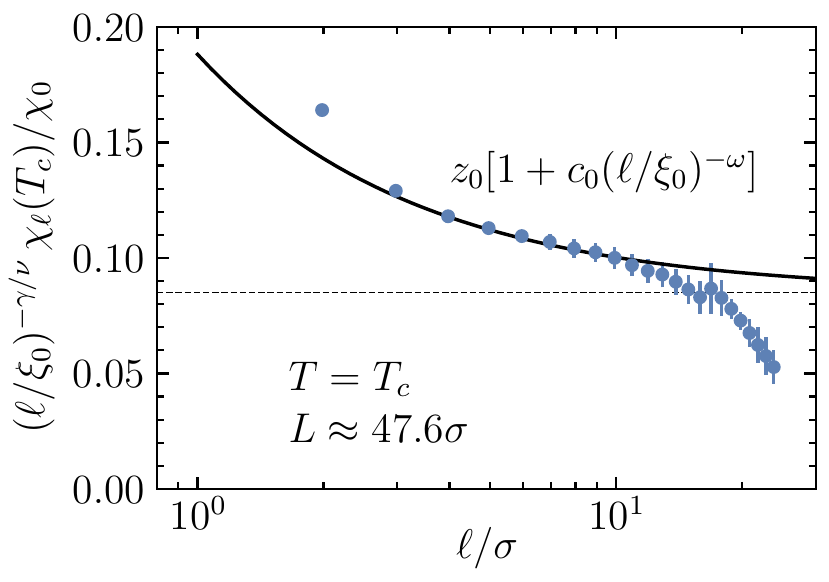}
 \caption{Confluent corrections to scaling of the critical susceptibility $\chi_\ell(T_c)$.
 Data were obtained at $T^*=1.421$ for a simulation box of $L\approx 47.6\sigma$ and were rectified by factoring out the leading power-law divergence.
 The solid line depicts the asymptotic law, \cref{eq:susceptibility_critical_scaling_corrections}, using
 $z_0 = 0.085$ and $c_0 = 2.8$ for the amplitudes,
 and the broken line indicates the value of $z_0$.
 }
 \label{fig:susc_corr_Tc}
\end{figure}

The situation is clarified by focussing on the behaviour at the critical point ($y=0$).
In this case, \cref{eq:susceptibility_corrections} implies for the susceptibility at $T=T_c$:
\begin{equation}
  \label{eq:susceptibility_critical_scaling_corrections}
  \chi_{\ell}(T_c) \simeq z_0 \chi_0 \, \left(\frac{\ell}{\xi_0}\right)^{\gamma/\nu}
    \Bigl[1+ c_0 (\ell/\xi_0)^{-\omega} \Bigr]\,,
  \quad \ell \gg \sigma,
\end{equation}
which is an extension of \cref{eq:susceptibility_critical_scaling};
the correction amplitude in dimensionless form is defined as $c_0 = \tilde c_0 \xi_0^{-\omega}$.
The equation suggests to divide the data for $\chi_\ell(T_c)$ by the critical divergence, $\chi_0 (\ell/\xi_0)^{-\gamma/\nu}$, so that the results would approach the constant $z_0$ for large $\ell$ and
the confluence is controlled by the term $\sim \ell^{-\omega}$.

\Cref{fig:susc_corr_Tc} shows such a plot of the simulation data for $\chi_\ell(T_c)$ as function of the sub-system size $\ell$,
with the rectified data actually calculated from the equivalent expression $a^2 \ell^{2\beta/\nu} \expect{\phi^2}_\ell$, see \cref{eq:op_scaling,fig:s2_s4_combined}.
However, the data do not converge for large $\ell$, rather they decay to zero.
This reveals a limitation of the sub-system analysis: the simulation box $L$ puts an upper limit on the accessible sub-system sizes $\ell$, and we anticipate deviations from \cref{eq:susceptibility_critical_scaling_corrections} unless $\sigma \ll \ell \ll L$.
Away from the critical point, this type of finite-size correction was shown to decay as $\ell^{-1}\e^{-\ell/\xi}$ [see Eq.~(30) in Ref.~\citenum{SSFslab:JCP2020}]; the ideas used there appear suitable to be transferred to the critical point, though the detailed analysis remains to be worked out.
Taking this issue into account, we have fitted \cref{eq:susceptibility_critical_scaling_corrections} to the data in \cref{fig:susc_corr_Tc} with only $z_0$ and $c_0$ as free parameters.
The fit was restricted to $3.5 \lesssim \ell/\sigma \lesssim 7.5$ and yielded $z_0 = 0.085(5)$ and $c_0 = 2.8(3)$,
where the errors were estimated from varying the bounds of the fit window by $\pm 0.5\sigma$.
Note that this value of $z_0$ is slightly smaller than the one obtained previously from fits to the temperature scaling of the 2\textsuperscript{nd} moment [\cref{fig:s2_s4_combined}].
The obtained curve is a reasonable description of the data and compatible with confluent corrections that scale as $\ell^{-\omega}$ at leading order.
In \cref{fig:susceptibility_collapse,fig:s2_s4_combined}, this type of correction appears to be of minor importance, it would be visible as deviations from the data collapse for small $\ell$ and arbitrary $\xi$.

\section{Binder cumulant for sub-systems}
\label{sec:binder_cumulant}

The data collapsing approach to determine $T_c$ as described in
\cref{subsec:chi_sub-system_scaling} has some subjective component.  A
superior method to locate a continuous phase transition was
established by \citet{Binder:1981ij,Binder:1981hj} and is based on the
4\textsuperscript{th} normalised cumulant $U_\ell(T)$ of the order
parameter distribution; it is closely related to the kurtosis used in
descriptive statistics.  Close to criticality, the composition
fluctuations $\phi=x_A-x_c$ are symmetric under sign change and one
defines the dimensionless ratio
\begin{equation}
  \label{eq:binder_cumulant}
 U_\ell(T) = 1 - \frac{\expect{\phi^4}_\ell}{3 \expect{\phi^2}_\ell^2} \,,
\end{equation}
where the subscript $\ell$, as before, denotes the linear extent of the
(sub-)system. At high temperature, the fluctuations are of Gaussian
nature, and thus $U_\ell(T \gg T_c) \to 0$.
Far below the critical temperature, the distribution has two sharp peaks at
the coexisting compositions, and the Binder cumulant tends to
$U_\ell(T \ll T_c) \to 2/3$.  As the critical temperature is
approached from either side of $T_c$, the correlation length in the
system diverges and the critical divergences of the moments cancel [cf.\ \cref{eq:op_scaling}]
so that $U_\ell(T_c)=: U_c$ remains finite and becomes independent of the sub-system size $\ell$.
The critical Binder cumulant $U_c = 1 - f_4(0) / 3 f_2(0)^2$
is a universal property of the
critical renormalisation group fixed point and as such depends only on the boundary conditions
and the geometric shape of the sub-system
\cite{Binder:1981ij,Binder:1981hj,Privman1984,Brezin1985,Kastening2013,Malakis:2014kh}.

The analysed open sub-systems mimick a grand canonical ensemble in the limit
of an infinitely large reservoir, $L \to \infty$, which is relaxed in
practice to the condition that the sub-system size~$\ell$ does not
compete with the finite extent~$L$ of the whole simulation box,
$\ell \ll L$; see ref.~[\onlinecite{SSFslab:JCP2020}] for a quantitative estimate.
We consider the idealised case $L \to \infty$ first,
corrections due to the finite simulation box are studied in \cref{sec:finite-size-corrections}.

\subsection{Common intersection point}
\label{subsec:binder_sub-system_scaling}

The basis of our discussion of the 4\textsuperscript{th} cumulant is the general scaling ansatz \cite{Binder:1981hj}
\begin{equation}
  \label{eq:binder_scaling_simple_ansatz}
  U_\ell(T) = \mathcal{U} \bigl(\ell^{1/\nu} \varepsilon \bigr) \,,
  \quad \varepsilon = (T-T_c)/T_c \,,
\end{equation}
employing a scaling function $\mathcal{U}(\cdot)$ that is analytic
since $U_\ell(T)$ describes finite (sub-)volumes and depends smoothly
on temperature.  Scaling is expected to hold when all length scales
are sufficiently large, in particular, when $\ell \gg \sigma$; the
ratio $\ell/\xi(T)$ is controlled by the scaling variable
$x:=\ell^{1/\nu} \varepsilon$.
Moreover, $\mathcal{U}(x)$ fulfills $\mathcal{U}(0) = U_c$ and
interpolates between the limits $\mathcal{U}(x\to \infty)=0$
and $\mathcal{U}(x\to -\infty)=2/3$.  Expanding
\cref{eq:binder_scaling_simple_ansatz} for small argument shows that,
near criticality, the quantity $U_\ell(T)$ varies linearly with
temperature around $U_c$:
\begin{equation}
  \label{eq:ul_near_Tc}
  U_\ell(T) \simeq U_c + u_1 \ell^{1/\nu} (T-T_c) \,,
  \quad T \to T_c,
\end{equation}
where $u_1:=\mathcal{U}'(0) \, T_c^{-1}$ is a non-universal constant.

\begin{figure*}
\centering
\includegraphics[width=.9\linewidth]{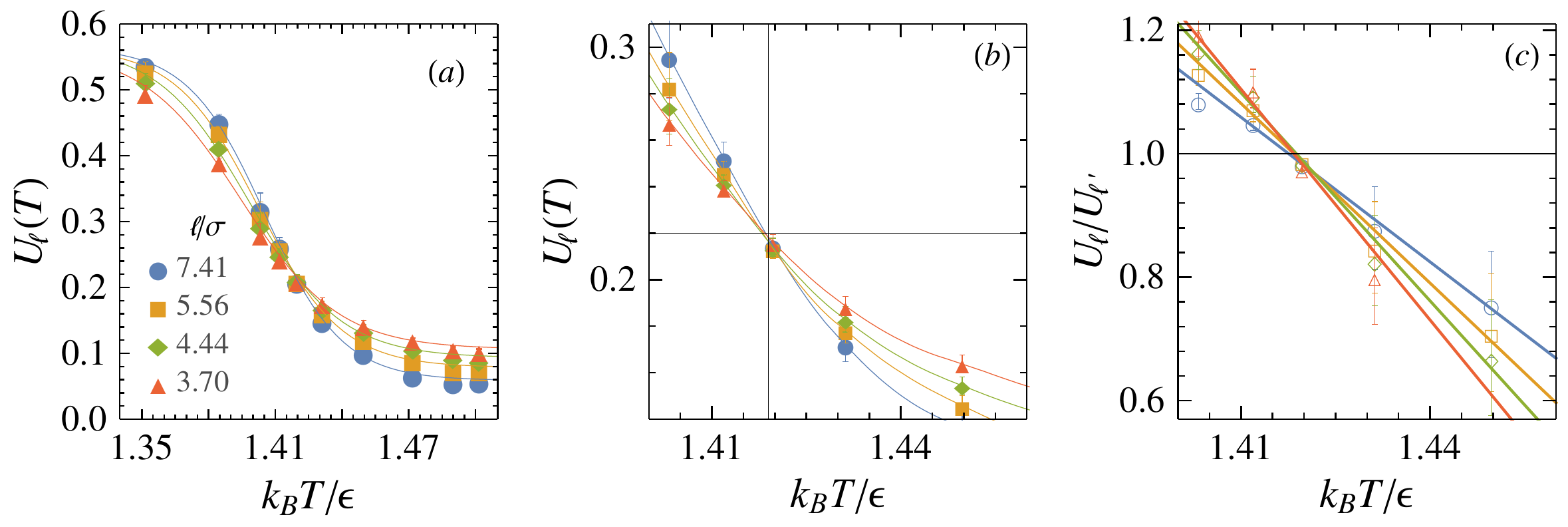}
\caption{(a) Simulation results for the Binder cumulant $U_\ell(T)$ for different
  sub-system sizes $\ell$ as indicated in the legend, with the same edge length
  $L \approx 44.4 \sigma$ of the overall simulation box.
  The data are based on the moments shown in \cref{fig:s2_s4_combined},
  and solid lines are fits of a $\tanh(\cdot)$-shape serving as a guide to the eye.
  ~~(b)~Close-up of the critical region of the same data as in panel~(a). Black lines indicate the common intersection point at the critical values $(T_c, U_c)$.
  ~~(c)~Cumulant ratios $U_\ell/U_{\ell'}$ as a function of temperature for different pairs of sub-system sizes:
  $(\ell,\ell')$ denoted by the symbols {\color{myblue} \large $\circ$}
  $(7.41\sigma,5.56\sigma)$, {\color{myokker} $\square$} $(7.41\sigma,4.44\sigma)$,
  {\color{mygreen} \large $\diamond$} $(7.41\sigma,3.70\sigma)$ and
  {\color{myorange} $\bigtriangleup$} $(7.41\sigma,3.18\sigma)$.
  Coloured lines are linear fits according to \cref{eq:cumulant_ratios}, and the horizontal black line marks the fixed point $U_\ell = U_{\ell'}$.}
  \label{fig:binder_cumulant}
\end{figure*}

Accordingly, a family of curves $U_\ell(T)$ for different sub-system
sizes $\sigma \ll \ell \ll \xi(T)$ has a common intersection point 
at $(T_c, U_c)$, in an asymptotic sense, which suggests a procedure to
locate the critical temperature $T_c$. It requires the simulation
of one large system for a number of temperatures sufficiently close to 
$T_c$, which in itself is challenging due to critical slowing down.
Our simulation results for sub-system sizes $\ell = L/m$ with $m=6,8,10,12$
and $L \approx 44.4 \sigma$ are shown in \cref{fig:binder_cumulant}(a,b).
The data exhibit a common intersection point as anticipated from \cref{eq:ul_near_Tc},
and we read off $T^{*}_c = 1.419$, in reasonable agreement with our previous result;
the critical value of the Binder cumulant $U_c = 0.22 \pm 0.01$ is close to
the earlier reported value in
Ref.~\citenum{Binder:1981ij} for Ising spins on a lattice.

\subsection{Scaling function}
\label{subsec:binder_scaling_function}

\begin{figure}
    \includegraphics[width=\figwidth]{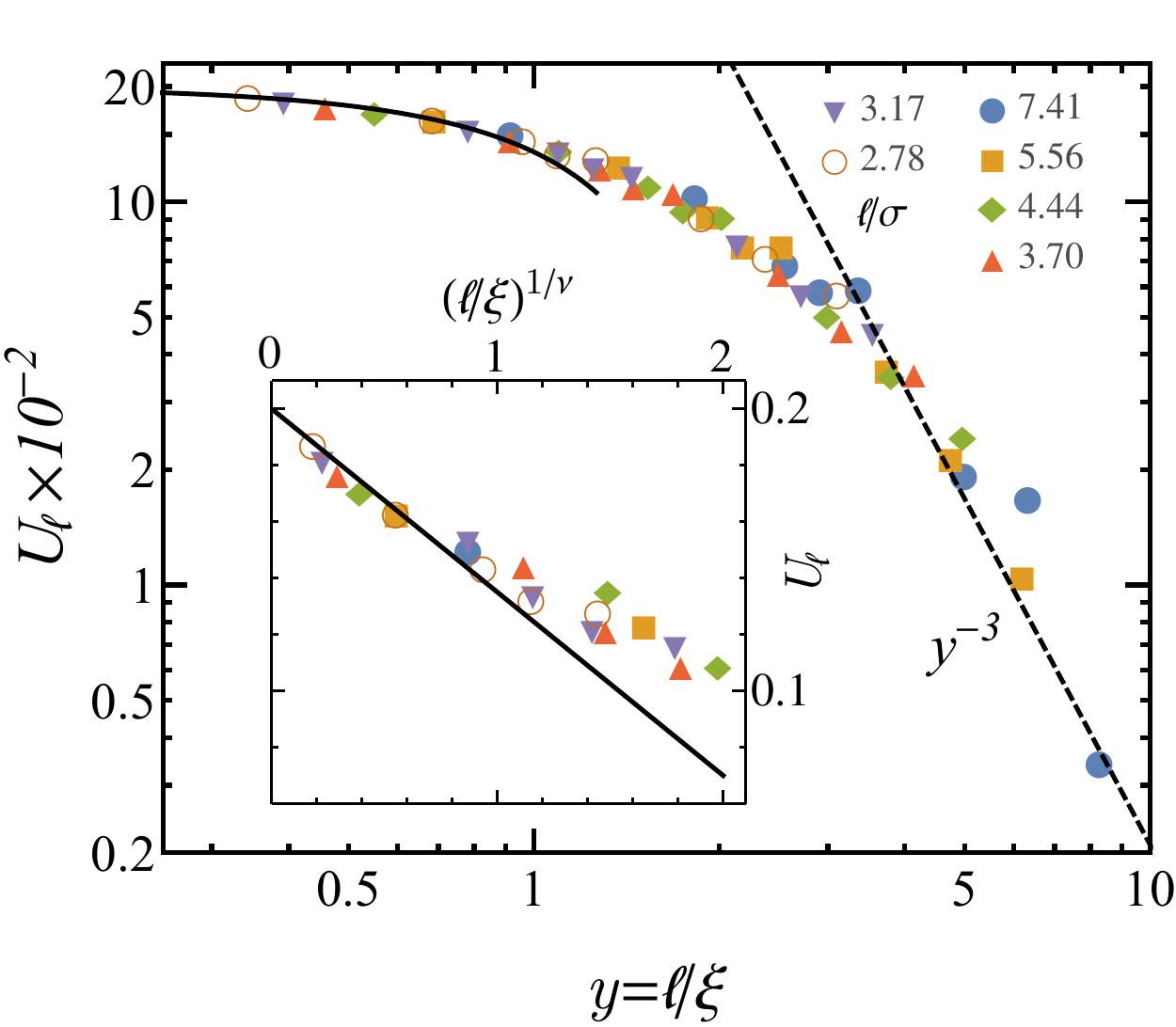}
    \caption{Test of the scaling form \cref{eq:binder_scaling_simple_ansatz}
      of the Binder cumlant $U_\ell(T)$ plotted against $y=\ell/\xi(T)$
      for the data in \cref{fig:binder_cumulant}
      but only for temperatures above $T_c$.  The dashed line
      indicates an algebraic decay $U_\ell(T) \sim [\ell/\xi(T)]^{-3}$.
      ~~Inset:
      Data collapse onto the scaling function $\mathcal{U}(x)$ is obtained by
      plotting the same data with $x=(\ell/\xi_0)^{1/\nu} \varepsilon$
      as the abscissa [\cref{eq:binder_scaling_simple_ansatz}].
      $\mathcal{U}(x)$ is analytic in $x=0$, and the linear extrapolation of the data
      [solid line, \cref{eq:ul_near_Tc}] intersects the vertical line $x=0$ at the universal value
      $U_c := U_\ell(T_c) \approx 0.20$.}
\label{fig:binder_cumulant_scaled_plot}
\end{figure}

A more physics-adapted way of writing the scaling form
\cref{eq:binder_scaling_simple_ansatz} for $T>T_c$ is
$U_\ell(T) = \hat{\mathcal{U}}_+(\ell/\xi)$ with
$\hat{\mathcal{U}}_+(y\geq 0) := \mathcal{U}\mleft((\xi_0
y)^{1/\nu}\mright)$,
noting that $\varepsilon \simeq (\xi/\xi_0)^{-1/\nu}$ in the critical
region.  Indeed, plotting our results for $U_\ell(T)$ against
$y=\ell/\xi(T)$ 
for different values of $\ell$
yields data collapse onto the scaling function
$\hat{\mathcal{U}}_+(y)$, see \cref{fig:binder_cumulant_scaled_plot}.
For small arguments, i.e., taking $T\to T_c$ for $\ell$ fixed, the
data converge to $\hat{\mathcal{U}}_+(y\to 0) = U_c$.  Near
$\ell/\xi \approx 1$, a crossover occurs to the high-temperature
regime (which implies small $\xi$, i.e., $\ell/\xi \to \infty$), where the
fluctuations are Gaussian and thus $\hat{\mathcal{U}}_+(y \to \infty) \to 0$.

What can be said about the convergence to the Gaussian regime, i.e.,
the asymptotics of $\hat{\mathcal{U}}_+(y \to \infty)$? An intuitive
argument assumes that the sub-volume of linear extent~$\ell$ can be
divided into independent ``correlation blobs'' of size~$\xi$ and
invokes a standard proof of the central limit theorem: consider the
sum $Y=\sum_{i=1}^n X_i$ of $n$ independent random variables $X_i$
that are identically distributed according to some characteristic
function $\varphi_X(\cdot)$ with variance $\sigma_X^2 < \infty$.  Then
$Y$ has the characteristic function
$\varphi_Y(\cdot) = \varphi_X(\cdot)^n$ and its cumulants $\mu_k$ are
generated by $n \log \varphi_X(\cdot)$, which shows that
$\mu_k \propto n$ for all $k=1,2,\dots$\,  The normalised variable
$Y/\sigma_Y$, with $\sigma_Y^2 := n \sigma_X^2$, becomes Gaussian as
$n \to \infty$ since its cumulants follow
$\mu_k \sigma_Y^{-k} \propto n^{1-k/2}$ and vanish for $k\geq 3$.

In the present context, we have to consider
$Y/n$ instead of $Y$ since the concentration $x_A$ in a sub-volume of
size $\ell$ is given by the arithmetic mean of the values of $x_A$ in
each of the $n=(\ell/\xi)^d$ correlation blobs. The cumulants of $Y/n$
are proportional to $n^{1-k}$, such that the normalised cumulant [\cref{eq:binder_cumulant}]
vanishes as $U_\ell \propto n^{-3} \big/ \mleft(n^{-1}\mright)^2 = n^{-1}$.
Thus, we expect that
$\hat{\mathcal{U}}_+(y) \sim y^{-d}$ for $y=\ell/\xi \to \infty$, here $d=3$.
Despite the limited availability of data for $U_\ell$ in this regime,
there is numerical evidence for the scaling $U_\ell \sim y^{-3}$ for $y \gtrsim 3$ as
shown in \cref{fig:binder_cumulant_scaled_plot}.
This scaling coincides with the behaviour derived for conventional finite-size scaling 
$U_L \sim L^{-d}$, see Eq.~(14) of Ref.~[\onlinecite{Binder:1981ij}]. 

Eventually, using $x=(\ell/\xi)^{1/\nu}$ as the scaling variable
rectifies the critical power-law
$\hat{\mathcal{U}}_+(y) - U_c \sim y^{1/\nu}$ for $y \to 0$
as a straight line, $\mathcal{U}(x) - U_c \sim x$ [\cref{eq:ul_near_Tc}],
which is supported by the inset of \cref{fig:binder_cumulant_scaled_plot}.

\subsection{Cumulant ratios}
\label{subsec:binder_cumulant_ratios}

An alternative method for estimating $T_c$ from $U_\ell(T)$ that does not depend
on the value of $U_c$ is to consider the ratio $U_\ell/U_{\ell'}$ as function of
temperature, where $\ell > \ell'$ are two different sub-system sizes.
For any choice of $\ell$ and $\ell'$ the ratio $U_\ell/U_{\ell'} = 1$ at $T = T_c$,
since $U_\ell(T)$ is independent of $\ell$ at the critical point,
see \cref{fig:binder_cumulant}(c). The behaviour for $T \to T_c$ follows from
\cref{eq:ul_near_Tc}
\begin{equation}
\label{eq:cumulant_ratios}
  \frac{U_\ell(T)}{U_{\ell'}(T)} \simeq 1
  + \frac{u_1}{U_c} \ell^{1/\nu} \mleft[1 - (\ell/\ell')^{-1/\nu}\mright](T-T_c) \,,
\end{equation}
which allows for a linear regression of the data for $U_\ell/U_{\ell'}$
near $T_c$ to find the intersection with unity.  A larger slope is
achieved for a larger sub-system size $\ell$ and a larger ratio
$\ell/\ell'$, whereas $\ell' \gg \sigma$ must not be chosen too small.
From this, we inferred $\kb T_c/\epsilon=1.4204 \pm 0.0008$, 
consistent with and slightly improving our
previous estimates of $T_c$ given above. 

\begin{figure}
\includegraphics[width=.9\linewidth]{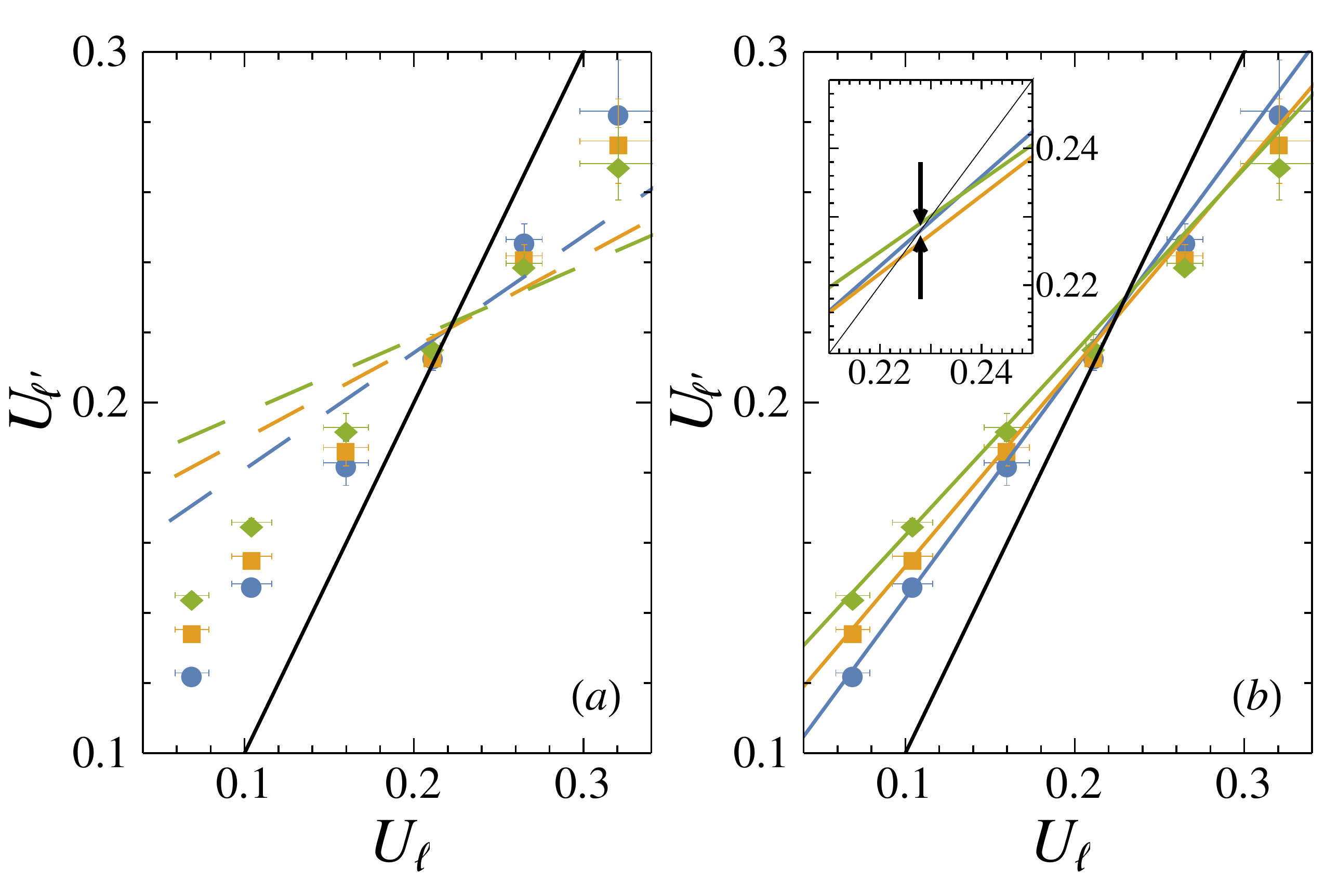}
\caption{(a) Graph of $U_{\ell'}(T)$ vs.\ $U_\ell(T)$ in the vicinity of the fixed point $(U_c, U_c)$ 
  for the investigated temperatures and for different pairs of the sub-system sizes
  $(\ell,\ell')$ with $\ell=7.41\sigma$ fixed and
  $\ell' = 5.56\sigma$ ({\color{myblue}\Large $\bullet$}),
  $4.44\sigma$ ({\color{myokker}\scriptsize $\blacksquare$}), and
  $3.18\sigma$ ({\color{mygreen}\scriptsize \protect\rotatebox[origin=c]{45}{$\blacksquare$}});
  the size of the simulation box was $L \approx 44.4\sigma$.
  Broken lines are fits of \cref{eq:ul_ul1_near_Tc} to the data points for the same $(\ell,\ell')$ with $U_c$ as only free parameter; the solid line indicates $U_\ell = U_{\ell'}$.
  ~~(b)~Same representation as in panel~(a) with solid lines showing linear fits to data with the slope and $U_c$ as parameters.
  The inset provides a close-up of the intersection of these lines with the diagonal $U_\ell=U_{\ell'}$;
  arrows indicate the range of an anticipated common intersection point.
  }
\label{fig:critical_binder_cumulant_finite_size}
\end{figure}

The determination of the critical Binder cumulant $U_c$ follows a
similar approach.  From the foregoing discussion it is clear that the
graph of $U_\ell(T)$ vs.\ $U_{\ell'}(T)$ for a given choice of $\ell$
and $\ell'$ displays a fixed point at $U_c$.  Thus, $U_c$ can be
determined from the intersection of this graph with the diagonal,
$U_\ell=U_{\ell'}$. Close to criticality, solving \cref{eq:ul_near_Tc}
for $T$ and substituting into $U_{\ell'}(T)$ yields a linear
relationship between $U_{\ell'}$ and $U_\ell$ (at the same
temperature):
\begin{equation}
\label{eq:ul_ul1_near_Tc}
  U_{\ell'} = U_c + (\ell/\ell')^{-1/\nu} \, (U_\ell - U_c) \,; \quad U_\ell \to U_c \,.
\end{equation}

Thus, $U_c$ will be the only free parameter in a linear regression
of the data for $(U_\ell, U_{\ell'})$.  The procedure is illustrated
in \cref{fig:critical_binder_cumulant_finite_size} with three
different choices of $(\ell,\ell')$.
The data for $U_\ell(T)$ plotted against $U_{\ell'}(T)$ 
for a range of temperatures $T$ and fixed ($\ell, \ell'$)
do indeed fall on straight lines as inferred from
\cref{eq:ul_near_Tc}.
However, the slopes do not match with the prediction $(\ell/\ell')^{-1/\nu}$ of \cref{eq:ul_ul1_near_Tc},
which points at a deficiency of the ansatz \cref{eq:binder_scaling_simple_ansatz}.
Nevertheless, permitting both $U_c$ and the slope as fit parameters yields approximately a common intersection point of the straight lines at $U_c \approx 0.22$ [\cref{fig:critical_binder_cumulant_finite_size}(b)].
A close-up of this intersection region reveals appreciable differences between the intersection points of two of the coloured curves (for different choices of $\ell'$) and their intersection with the diagonal.
The failure of \cref{eq:ul_ul1_near_Tc} and this observation led us to revisit our scaling ansatz
and to rederive \cref{eq:ul_near_Tc} in the next section. The error stems from the coefficient $u_1$, which was
obtained as $u_1:=\mathcal{U}'(0)T_c^{-1}$, that is, as constant with respect to~$\ell$.
However, taking into consideration the finite size of the
simulation box shows that in fact $u_1$ depends on $\ell/L$.

\subsection{Finite-size corrections}
\label{sec:finite-size-corrections}

So far, we have ignored the finiteness of the total simulation volume,
which can taint the estimates of the critical point, including both
$T_c$ and $U_c$.  In particular, there is a competition between the
sub-system size $\ell$ and the length $L$ of the whole simulation box,
leading to a kind of effective boundary conditions on the
sub-system as $\ell$ grows. (For example, consider $\ell = L/2$, which implies boundary conditions that are neither free nor periodic.)
Further, the correlation length $\xi$ enters as a third length scale, and it
may be necessary to distinguish the regimes $\ell \ll \xi \ll L$ and
$\ell \ll L \ll \xi$, in addition to $\xi \ll \ell, L$.  In the
following, we will assess the importance of these corrections for our
results combining theoretical arguments and simulation data for a
range of box sizes~$L$.

\begin{figure}
  \centering
  \includegraphics[width=\figwidth]{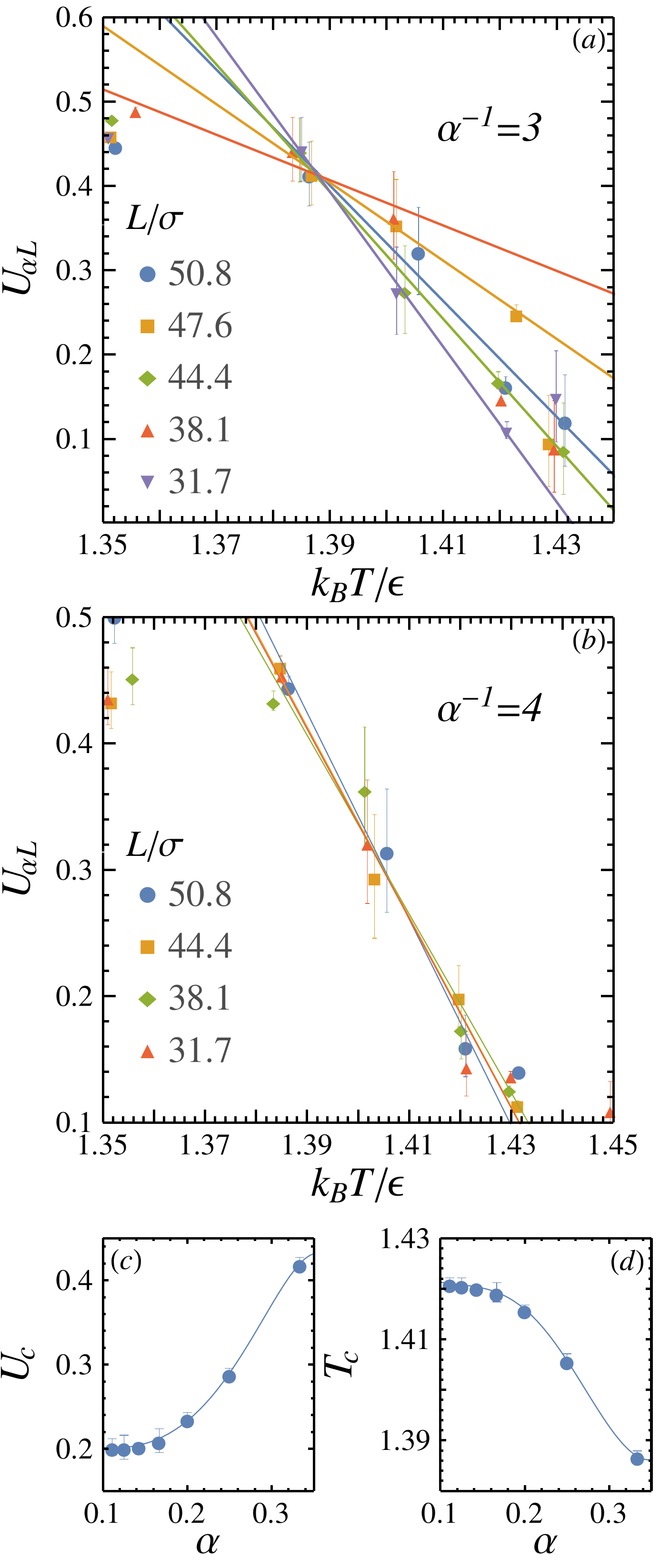}
  \caption{
    (a),(b)~Binder cumulant $U_\ell(T;L)$ as a function of temperature
    for fixed ratios $\alpha=\ell/L$ between the sizes of the 
    sub-system ($\ell$) and the simulation box ($L$) as indicated in each panel.
    Linear fits (solid lines) in the critical region exhibit a common intersection point 
    at $\bigl(\tilde U_c(\alpha), T_c(\alpha)\bigr)$, which determines
    effective, $\alpha$-dependent values for $U_c$ and $T_c$ according to \cref{eq:ul_near_Tc_finite}.
    ~~(c),(d)~Results for $U_c(\alpha)$ and $T_c(\alpha)$, respectively, obtained from
    aspect ratios of $\alpha^{-1}=3,4,\dots,9$ for system sizes ranging from $L\approx 31.7\sigma$ to $50.8\sigma$.
    Solid lines are smooth interpolations of the data to guide the eye.
  }
\label{fig:finite_size_corrections}
\end{figure}

Conventional finite-size scaling is based on the fact that near $T_c$,
the correlation length exceeds the system size by far.  For the
sub-system analysis, the corresponding regime is $\ell, L \ll \xi$ and
we expect that the predominant corrections due to finite $L$ are
controlled by the aspect ratio $\alpha := \ell/L$.
This suggests to
amend the scaling ansatz \eqref{eq:binder_scaling_simple_ansatz} by
the variable $\ell/L$ to
\begin{equation}
  \label{eq:binder_scaling_ansatz}
  U_\ell(T; L) = \tilde{\mathcal{U}} \mleft(\ell^{1/\nu} \varepsilon, \ell / L \mright) \,,
  \quad \varepsilon = (T-T_c)/T_c \,,
\end{equation}
which is expected to hold when all length scales are sufficiently
large, i.e., for $\ell, L, \xi(T) \gg \sigma$.
(An alternative to this ansatz is discussed briefly in \cref{sec:appendix_a}.)
Taking $L\to \infty$ yields \cref{eq:binder_scaling_simple_ansatz},
and in terms of the scaling functions:
$\mathcal{U}(x) = \tilde{\mathcal{U}}(x, \alpha \to 0)$.  The function
$\tilde{\mathcal{U}}(x,\alpha)$ is analytic in $x=0$ for any fixed
$\alpha \geq 0$, which we infer from the fact, used before, that in a
finite system thermodynamic observables depend smoothly on
temperature.  The dependence on $\alpha$, on the other hand, is not
known albeit an exponentially fast approach to the thermodynamic limit
is not unlikely:
$\tilde{\mathcal{U}}(x,\alpha) - \tilde{\mathcal{U}}(x, 0) =
O\bigl(\e^{-1/\alpha}\bigr)$ as $\alpha \to 0$.

Expanding \cref{eq:binder_scaling_ansatz} around $x=0$ yields close to criticality:
\begin{equation}
  \label{eq:ul_near_Tc_finite}
  U_\ell(T; L) \simeq \tilde U_c(\ell/L)  + \tilde u_1(\ell/L)\, \ell^{1/\nu} (T-T_c) \,,
\end{equation}
as $T \to T_c$,
where we introduced an effective critical cumulant as
$\tilde U_c(\alpha) := \tilde{\mathcal{U}}(0, \alpha)$ and the
geometry-dependent coefficient
$\tilde u_1(\alpha) := T_c^{-1} \partial_x \tilde{\mathcal{U}}(x,
\alpha) \big|_{x = 0}$.
It becomes evident from \cref{eq:ul_near_Tc_finite} that the curves
$U_\ell(T;L)$ vs. $T$ for different $\ell$, but the same system size
$L$, do not have a common intersection point.  Such a point emerges
only asymptotically for $\ell/L$ sufficiently small such that
$\tilde U_c$ and $\tilde u_1$ do not depend appreciably on $\ell/L$,
i.e., in the limit $L\to \infty$.

Nevertheless, a common intersection point at
$(T_c, \tilde U_c(\ell/L))$ is achieved for fixed ratios
$\alpha=\ell/L$, since then the $\ell$-dependence enters only the term
proportional to $T-T_c$ in \cref{eq:ul_near_Tc_finite}.  In practice,
the reciprocal $m=1/\alpha$ is an integer number counting the sub-volumes
(along each axis) that fit into the whole system.  Thus, the refined
analysis procedure accounting for finite system sizes $L < \infty$
would be as follows: From a set of simulations for a few system sizes
$L$ and many $T$ values in the critical regime, compute the sub-system
statistics and $U_\ell(T;L)$ in particular for selected values of
$\alpha$. Data for different $\alpha$ are analysed separately.  For
given $\alpha$, plotting $U_{\alpha L}(T;L) / U_{\alpha L'}(T;L')$ as
function of $T$ the value of $T_c(\ell/L)$ can be read off from the
intersection with unity; more precisely, it follows from the linear
regression according to [cf.\ \cref{eq:cumulant_ratios}]
\begin{multline}
\label{eq:cumulant_ratio_finite_size}
  \frac{U_{\alpha L}(T; L)}{U_{\alpha L'}(T; L')} \simeq \\
  1 + \frac{\tilde u_1(\alpha)}{U_c(\alpha)} \, (\alpha L)^{1/\nu} \, \mleft[1 - (L'/L)^{1/\nu}\mright] (T - T_c)
\end{multline}
in the limit $T \to T_c$.

Since \cref{eq:cumulant_ratio_finite_size} follows directly from
\cref{eq:ul_near_Tc_finite}, we test this finite size scaling analysis
on the measured data for the Binder cumulant. To this end, we consider
$U_{\alpha L}(T)$ as a function of temperature for
different combinations of $\ell$ and $L$ but with $\alpha$ fixed. 
We have carried out additional simulations with different box lengths
varying from $L \approx 31.7 \sigma$ ($32{\,}000$ particles)
to $L \approx 50.8 \sigma$ ($131{\,}072$ particles),
and the results of the analysis are shown in \cref{fig:finite_size_corrections}.

Since $T_c(\alpha)$ and $U_c(\alpha)$ are \emph{a priori} not known, we fitted
the data $U_{\alpha L}$ near the critical temperature to a linear
function [\cref{eq:ul_near_Tc_finite}].
For each value of $\alpha$, this lines exhibit a well-defined
common intersection point, from which we have read off
$T_c(\alpha)$ and $U_c(\alpha)$ [\cref{fig:finite_size_corrections}(a,b)].
We note that for $\alpha^{-1} \geq 10$, no common intersection point 
could be detected due to almost equal slopes, why we restricted the analysis to 
$\alpha^{-1}=3,4,\dots,9$.
The values obtained for $U_c(\alpha)$ and $T_c(\alpha)$
show a considerable dependence on $\alpha$, yet they converge
as $\alpha \to 0$ [\cref{fig:finite_size_corrections}(c,d)].
An extrapolation of the data to this limit, i.e., for $L \gg \ell$, yields
our final estimates of \emph{(i)}~the universal value for the critical Binder cumulant $U_c(0)=0.201 \pm 0.001$
and \emph{(ii)}~the critical temperature $T_c=1.421 \pm 0.001$ specific to the investigated binary liquid.

\section{Summary and conclusions}
\label{sec:conclusion}

We have given a comprehensive analysis of the local order parameter fluctuations in open sub-systems
and compared different approaches to locate the critical temperature $T_c$.
The applicability of the procedures was demonstrated for a symmetric binary liquid with known phase diagram, fully based on molecular dynamics simulations of one large system.
Thus, such simulations provide an alternative to 
(semi-)grand canonical Monte Carlo schemes, and arguably have the advantage of simultaneously probing the dynamic properties of the system, e.g., transport coefficients (which we have not discussed here).
Complementary, the study of the static structure factor calculated for a large system size yields the critical divergences of the correlation length and the susceptibility (\cref{fig:scck}), which were found to be compatible with the 3D-Ising universality class as expected and from which we got a first estimate of $T_c$.

For the composition fluctuations, obtained from the particle number statistics in cubic sub-volumes, we have shown that the standard finite-size scaling procedures, as for a sequence of simulations with periodic boundaries, are successful if the edge length $L$ of the simulation box is replaced by that of the of sub-volume ($\ell$).
In particular, the data for the susceptibility $\chi_\ell(T)$ and the 4\textsuperscript{th} moment $\expect{\phi^4}_\ell$ collapse onto master curves after appropriate rescaling [see \cref{eq:susceptibility_scaling,eq:op_scaling}] and if $T_c$ is chosen properly (\cref{fig:susceptibility_collapse,fig:s2_s4_combined}).
Confluent corrections to scaling are compatible with a decay $\sim \ell^{-\omega}$ with $\omega\approx 0.83$ and are practically relevant due to computational limitations on exploring the ideal scaling regime $\sigma \ll \ell \ll L$.
For example, the susceptibility at $T=T_c$ deviates from its critical power-law divergence by still 20\%
for a sub-system size of $\ell = 10\sigma$, which corresponds to $\ell \approx L/5$ for the comparably large simulation box used here (\cref{fig:susc_corr_Tc}).

Further, we have found that the Binder cumulant $U_\ell(T)$ of the sub-systems, plotted against temperature $T$, yields only an apparent common intersection point (\cref{fig:binder_cumulant}).
Nevertheless, it yields $T_c$ to an accuracy of about 0.2\% in our example,
where we partitioned the simulation box into $m^3$ cubes for $m=6,\dots,12$ with $\ell=3.7\sigma$ for the smallest sub-systems.
Here, it was favourable to consider the cumulant ratios $U_\ell/U_{\ell'}$, which does not require knowledge of the critical value $U_c$.
Due to the free boundary conditions, the latter adopts a universal value \cite{Binder:1981ij} $U_c \approx 0.2$ that is very different from its 3D-Ising value for periodic boundaries ($U_c^\mathrm{per}\approx 0.624$).

Extending the finite-size scaling ansatz [\cref{eq:binder_scaling_simple_ansatz}] for $U_\ell(T)$ by the aspect ratio $\alpha=\ell/L$ as a second scaling variable, we have shown that a true common intersection point is predicted asymptotically and observed in our simulation data, provided that lines of constant $\ell/L$ are considered [\cref{eq:ul_near_Tc_finite} and \cref{fig:finite_size_corrections}].
A disadvantage of this more correct approach is that it requires again a sequence of simulations for a number of large system sizes, so that $\ell$ can be varied while keeping $\ell/L$ fixed.
Despite the existence of a common intersection point for aspect ratios even close to unity (e.g., $\alpha = 1/3$), we showed that its location $(U_c, T_c)$ can sensitively depend on $\alpha$, but converged for $\alpha \lesssim 0.1$ in our case.
A similarly large aspect ratio between sub-system and overall simulation box is needed to establish free boundary conditions on the surfaces of the sub-system as derived by one of us recently \cite{SSFslab:JCP2020}.
From the extrapolation $\alpha \to 0$, we improved previous estimates of the critical Binder cumulant to $U_c = 0.201 \pm 0.001$; the value applies for physical systems in the short-ranged 3D-Ising universality class and if cubic domains with free boundaries are considered.
Interestingly, our estimates for the critical temperature, $T_c = 1.421 \pm 0.001$, did not change appreciably for the various approaches used here and are in agreement with previously reported values for this particular model system \cite{Das2006, Roy:2011jl}.

In conclusion, we have shown that the analysis of open sub-systems offers a reliable method to locate critical points, thereby taking advantage of large-scale simulations facilitated by massively parallel computing hardware.
A complication arises due to the interference of the sub-system size $\ell$ with the size $L$ of the simulation box, which requires a finite-size scaling procedure with $\ell/L$ fixed;
the latter defeats the idea of sticking to a single, large value of $L$.
Only for $\ell/L \lesssim 1 / 10$ or even less, free boundary conditions are effectively realised on the sub-volume surfaces and the dependence on $L$ drops out.
In many practical situations, this allows resorting to the simplified analysis with $L$ fixed [\cref{fig:binder_cumulant}], i.e., based on one or few simulation runs of one large system.
On the other end of the scale, we found that sub-volume sizes as small as $\ell \approx 3{-}4\sigma$ still permit scaling, so that the choice $L\approx 50{-}100 \sigma$ yields sufficient room for varying the sub-system size $\ell$.
A spin-off from a large ratio $m=L/\ell$ is that the simulation data at a single instance in time permit a statistical average over $m^3$ sub-systems, similarly as contributions at small wavenumbers $k$ to the structure factor $S_\mathrm{cc}(k)$ are self-averaging.
Eventually, the presented sub-system analysis combined with the two-parameter scaling should also be applicable to the more challenging study of asymmetric, e.g., colloid--polymer mixtures \cite{Liu:PRL2006,Zausch:JCP2009,Trefz:JCP2017}.
Apart from specific applications to fluids and taking advantage of universality, Monte Carlo simulations of large 3D Ising lattices should provide a highly sensitive test of the extended finite-size scaling of $U_\ell(T;L)$ [\cref{eq:ul_near_Tc_finite,eq:cumulant_ratio_finite_size}] and a more precise estimate of $U_c$.

\appendix

\begin{acknowledgments}
  We thank Siegfried Dietrich (MPI-IS Stuttgart) and Surajit Sengupta (TIRF Hyderabad) for useful discussions.
  This research has been supported by Deutsche Forschungsgemeinschaft (DFG) through grant SFB~1114, project no.\ 235221301, sub-project C01.
  DC acknowledges funding from BRNS vide grant number
  37(3)/14/10/2018-BRNS/370132 and YP acknowledges funding from DST
  Women Scientist Scheme via grant number SR/WOS-A/PM-36/2017(G). 
\end{acknowledgments}

\section{Alternative scaling ansatz for $U_\ell(T;L)$}
\label{sec:appendix_a}

It is tempting to propose as a natural extension of the scaling ansatz
\cref{eq:binder_scaling_simple_ansatz} that
\begin{equation}
  \label{eq:binder_scaling_ansatz_alt}
  U_\ell(T; L) = \dbltilde{\mathcal{U}} \mleft(\ell^{1/\nu} \varepsilon, L^{1/\nu} \varepsilon\mright) \,,
  \quad \varepsilon = (T-T_c)/T_c \,,
\end{equation}
which is expected to hold when all length scales are sufficiently
large, i.e., for $\ell, L, \xi(T) \gg \sigma$.  Taking $L\to \infty$
yields \cref{eq:binder_scaling_simple_ansatz}, and in terms of the
scaling functions,
$\mathcal{U}(x) = \dbltilde{\mathcal{U}}(x, X \to \infty)$ with
$X:= L^{1/\nu} \varepsilon$, so that
$\lim_{x\to 0} \dbltilde{\mathcal{U}}(x, X \to \infty) = U_c \approx
0.20$
recovers the universal Binder cumulant for open boundaries.  On the
other hand, choosing $\ell = L$ reproduces the conventional
finite-size scaling with periodic boundaries
\cite{Bloete1995,Bloete1999}, and so
$\lim_{x \to 0} \dbltilde{\mathcal{U}}(x, x) = U_c^\mathrm{per} \approx
0.624$.
From a practical perspective, effective values of $U_c$ are obtained
from carrying out the data analysis described in
\cref{subsec:binder_sub-system_scaling} for various finite~$L$.
We thus define
$U_c(\alpha) := \lim_{x \to 0} \dbltilde{\mathcal{U}}(x, \alpha^{-1/\nu}
x)$
where $\alpha := \ell / L$ fixes the aspect ratio of the geometry.
With this, the limit $x\to 0$ becomes a function
of $\alpha$ and interpolates between $U_c$ for $\alpha = 0$ (free
boundaries) and $U_c^\mathrm{per}$ for $\alpha = 1$ (periodic
boundaries).  In particular, the scaling function
$\dbltilde{\mathcal{U}}(x, X)$ is not continuous at the origin,
$x = X = 0$, not even speaking of analyticity. Yet, the function is
analytic in $x$ alone for any fixed $L \leq \infty$ since $U_\ell(T;L)$
depends smoothly on temperature in a finite (sub-)system.  The
standard treatment to derive asymptotic scaling behaviour, however,
fails as it relies on a Taylor expansion of the scaling function
around the critical point ($\varepsilon = 0$), and at the bottom line
the ansatz \eqref{eq:binder_scaling_ansatz_alt} proves fruitless.

\bibliography{critical_dynamics}

\end{document}